\documentclass[
	a4paper, % Paper size, use either a4paper or letterpaper
	10pt, % Default font size, can also use 11pt or 12pt, although this is not recommended
	twoside, % Two side traditional mode where headers and footers change between odd and even pages, comment this option to make them fixed
]{LTJournalArticle}

\runninghead{An Evaluation of Immersive Infographics for News Reporting}
\footertext{\textit{This is preprint version of a manuscript submitted to peer review.}} 

\usepackage{mathptmx}                  % use matching math font
\usepackage{bm}
\usepackage{wrapfig}
\usepackage{graphicx,subfigure}

\newcommand\trash{\includegraphics[width=0.12in,height=0.12in]{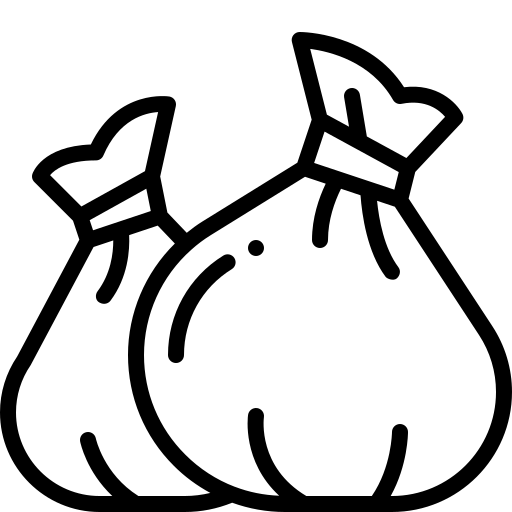}}
\newcommand\water{\includegraphics[width=0.12in,height=0.12in]{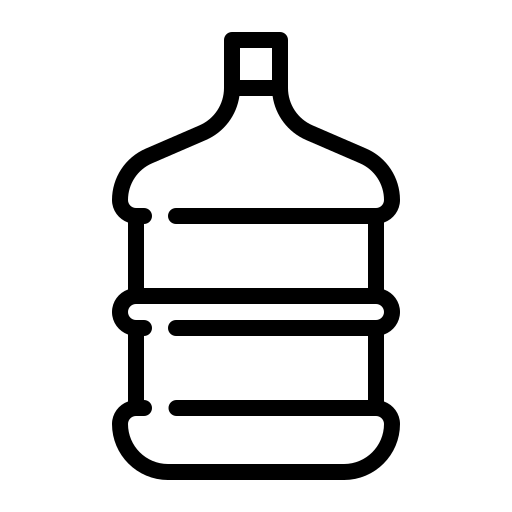}}
\newcommand\money{\includegraphics[width=0.12in,height=0.12in]{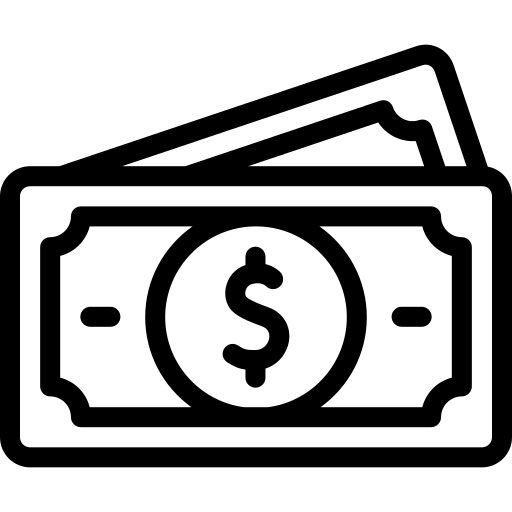}}
\usepackage[]{mdframed}
\usepackage{xurl}
\usepackage{pifont}% http://ctan.org/pkg/pifont
\newcommand{\cmark}{\ding{51}}
\newcommand{\xmark}{\ding{55}}

\usepackage{hyperref}

\renewcommand\footnotemark{}

\title{An Evaluation of Immersive Infographics for News Reporting: Quantifying the Effect of Mobile AR Concrete Scales Infographics on Volume Understanding} 
\author{%
	Mariane Giambastiani, Jorge Wagner, Carla M. Dal Sasso Freitas and Luciana Nedel\thanks{\textit{This is preprint version of a manuscript submitted to peer review.} Corresponding author: \href{mailto:marianegiamb@gmail.com}{marianegiamb@gmail.com}}%\textbf{Received:} October 20, 2023, \textbf{Published:} December 14, 2023}
}

\date{\footnotesize Institute of Informatics, Federal University of Rio Grande do Sul, Brazil}

\renewcommand{\maketitlehookd}{%
	\begin{abstract}
\noindent Augmented Reality (AR) allows us to represent information in the user's own environment and, therefore, convey a visceral feeling of its true physical scale. Journalists increasingly leverage this opportunity through immersive infographics, an extension of conventional infographics reliant on familiar references to convey volumes, heights, weights, and sizes. Our goal is to measure the contribution of immersive mobile AR concrete scales infographics to the user's understanding of the information scale. We focus on infographics powered by tablet-based mobile AR, given its current much more widespread use for news consumption compared to headset-based AR. We designed and implemented a study apparatus containing three alternative representation methods (textual analogies, image infographic, and AR infographic) for three different pieces of news with different characteristics and scales. In a controlled user study, we asked 26 participants to represent the expected volume of the information in the real world with the help of an AR mobile application. We also compared their subjective feelings when interacting with the different representations. 
While both image and AR infographics led to significantly better comprehension than textual analogies alone across different kinds of news, AR infographics led, on average, to a 31.8\% smaller volume estimation error than static ones. 
Our findings indicate that mobile AR concrete scales infographics can contribute to news reporting by increasing readers' abilities to comprehend volume information.
\end{abstract}
}

\begin{document}

\maketitle % Output the title section

%----------------------------------------------------------------------------------------
%	ARTICLE CONTENTS
%----------------------------------------------------------------------------------------

\section{Introduction}
\label{sec:intro}

When reading the news, either on paper or through a news app or website, we often find infographics, i.e., visualizations enriched with illustrations and annotations to present information \cite{burns2022pictographs}.
 \textit{Immersive infographics} leverage augmented (AR) or virtual (VR) realities to increase information understanding using infographic narratives to offer an alternative perspective, comprising, in our view, one of the most exciting applications of \textit{Immersive Analytics}
\cite{immersiveanalyticsbook}. In particular, through infographics based on the concept of \textit{concrete scales} \cite{Chevalier_2013}, we can provide users with the sense of \textit{data visceralization} \cite{Lee_2020}, conveying volumes or dimensions that can often be difficult to estimate through abstract numbers alone.

\begin{figure*}
\includegraphics[width=\linewidth]{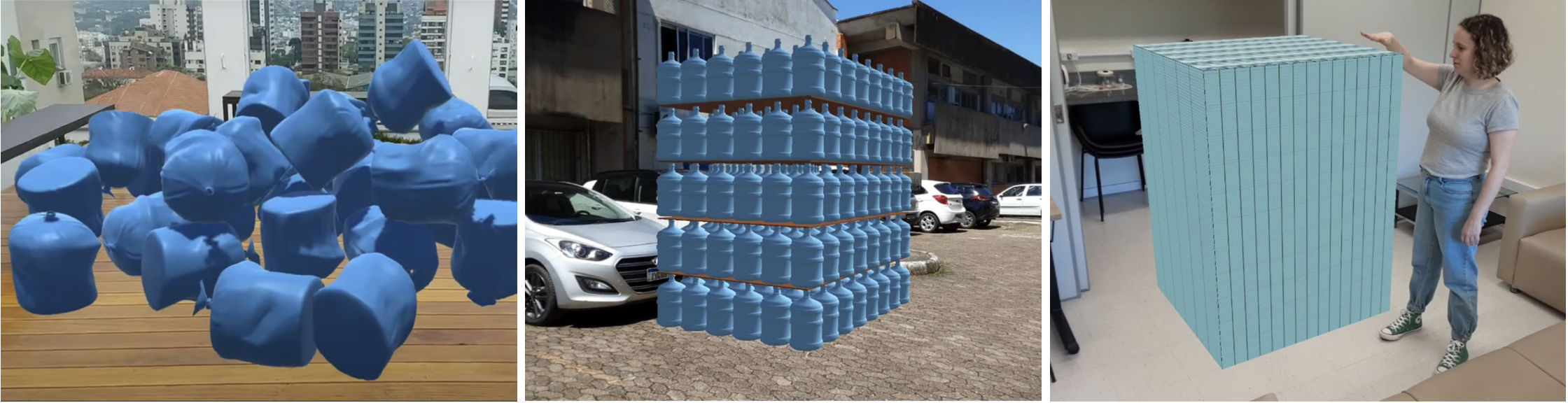}
\caption{
Mobile news apps, widely used for news consumption, can leverage Augmented Reality to contextualize volume information in news articles in relation to the reader's real environment. This can potentially improve understanding of information on waste production, water use, and amounts of money, as we have proposed. Our goal is to compare the user's comprehension of volumes expressed in AR against two more conventional approaches: textual analogies and static images of concrete scales infographics.}
\label{fig:teaser}
\end{figure*}

Not surprisingly, media companies have been investing in this kind of application as a way of improving their readers' experiences, as exemplified, most prominently, by The New York Times (NYT) \cite{nyt2022mixedrealityfuture}. The NYT started exploring AR reporting in 2018 \cite{nyt2018ar} and has since published at least 31 examples of ``AR storytelling'' \cite{nyt2022examples}, including many that could be classified as \textit{immersive infographics}. Most of them have been produced in partnership with Instagram and work through the popular Instagram application as an ``AR filter'' \cite{NYT_Angela_2020}. They complement news articles where conventional and animated infographics are also presented, offering the opportunity for a different perspective. The news company has also stated the intention of targeting mixed reality headsets in the future \cite{nyt2022mixedrealityfuture}. 

Some of the AR infographics presented by the NYT offer a visceral understanding, for example, of the meaning of a given amount of rain measured in inches \cite{nyt2022CaliforniaMegastorm} or of the performance of Olympic athletes \cite{nyt2022RaceGoldMedalist}. Some leverage a first-person egocentric perspective, for example, to illustrate how virus particles flow in the air inside a classroom under different ventilation conditions \cite{nyt2022ReopeningSchools}. Other examples also demonstrate how immersive 3D infographics, even in AR, can help explain very small or large phenomena, such as how face masks intercept virus particles (in a much enlarged 3D representation) \cite{nyt2022HowMasksWork} or how wildfires can cause storms (in a miniature 3D representation) \cite{nyt2022WildfireStorms}.

In this work, after reviewing background concepts and related works (Section \ref{sec:background}), we report on a controlled user study designed to measure the effect of immersive infographics on volume understanding (Section \ref{sec:experiment}). For this study, we focus specifically on concrete scales immersive infographics implemented through a mobile AR tablet application, simulating a 3D news visualization (see Figure \ref{fig:teaser}). Our central hypothesis is that such kind of concrete scales immersive infographics should help people understand large numerical quantities better than a conventional static infographic in image format without moving from a tablet to a different device---e.g., a see-through head-mounted display.

To investigate this hypothesis, in our comparative study, we asked 26 participants to indicate the volume they imagined when reading news about large volumes of trash, money, and water through an AR volume specification tool. Additionally, we used a \textit{Smile or Scowl} assessment \cite{Lan_2021} to measure how participants felt when reading the news in three different viewing modes: text, image, and AR. 

In summary, our work contributes with:
\begin{itemize}
\item A controlled user study designed to investigate the potential benefits of immersive concrete scales infographics in volume understanding tasks.
\item Findings indicating that immersive concrete scales infographics make it easier to understand large quantities than static image-based concrete scales infographics. 
\end{itemize}

\section{Related Work}
\label{sec:background}

Our work explores the use of well-known concrete scales representations (Section \ref{sec:rel:concrete_scales}) in combination with immersive infographics (Section \ref{sec:rel:immersive_ifographics}), a novel form of storytelling that is becoming more common with recent advances in AR devices (both mobile and head-mounted). While some prior work has already delved into the possibilities enabled by immersive concrete scales representations and the related concept of \textit{data visceralization} (Section \ref{sec:rel:imm_conc_info}), to the best of our knowledge, there is limited information available on the scale of the effect of such representations on user volume understanding, which is the gap we seek to fill with our controlled study. Below, we briefly survey the most closely related literature on these concepts to better contextualize our investigation.
\color{black}

\subsection{The Use of Concrete Scales Representations}
\label{sec:rel:concrete_scales}

News stories often resort to analogies, either in textual or graphical format, as a resource to more efficiently convey complex numeric information, putting it in perspective to help readers understand and remember data they could have trouble picturing in its original format \cite{hullman2018concrete,riederer2018analogies}. Using the concept of \textit{concrete scales}, i.e., visually re-expressing the information in terms of magnitudes and units that are easier to grasp, as defined by Chevalier et al. \cite{Chevalier_2013}, is particularly helpful in informing quantities and dimensions. Concrete scales can help readers understand, for example,  how enormous quantities of money compare to the ones they are used to in their personal lives. Chevalier et al. exemplified this with a visualization of illustrative stacks of banknotes comparing their sizes to landmark New York City buildings. 
A similar example was later revisited by Lee et al. in the context of VR \cite{Lee_2020}.

Concrete scales infographics can be expressed as 3D images that convey the information as a narrative, using terms and units that one can easily understand. The Democracy.info \cite{democracy_info} website, for example, presents a set of concrete scales infographics. The infographic \textit{All the gold in the world} \cite{democracy_info_Gold} shows us the size of all gold already mined in the world as if it were molten into one solid cube. Meanwhile, \textit{US Debt Visualized in} \$\textit{100 Bills} \cite{democracy_info_US_debt} represents how much space would be needed to store 30 trillion dollars (the debt as of 2022) in \$100 bills.

Concrete scales representations can also be employed in a physical format. Lindrup et al. \cite{Lindrup2023Physicalization}, for example, used data physicalizations to depict the amount of carbon emitted in the food manufacturing process. They chose three types of food, asparagus, cheese, and hamburger, and displayed the amounts of carbon emitted in each part of their manufacturing as wooden blocks.

\subsection{\scalebox{.95}[1.0]{The Increasing Adoption of Immersive Infographics}}
\label{sec:rel:immersive_ifographics}

Most of us are probably used to seeing infographics in printed format or even in online news \cite{Provvidenza_2019,Lu2020Infographics}. With the widespread adoption of mobile devices equipped with AR capabilities and the growing use of mobile applications for news consumption---according to Pew Research, 86\% of Americans get news on digital devices \cite{pew2021digital}---, we are now also able to visualize infographics in 3D space, what we refer to as \textit{immersive infographics}.

Recent works have already mentioned the use of infographics in immersive settings. 
Isenberg et al. \cite{Isenberg_2018}, for example, discussed the challenges and opportunities for immersive visual data stories. One of the possibilities they pointed out to improve knowledge about data was the use of infographics for immersive storytelling. In this case, the immersive approach did not use VR/AR but storytelling techniques to involve the reader in the infographic story's context. 
On the other hand, Zhu et al. \cite{Zhu_2024} used VR for immersive data storytelling to inform people about health situation awareness. They presented to the participants how a virus such as the one responsible for COVID-19 spreads through the air in a closed room with closed or open windows to educate people about health threats. They pointed out that immersive storytelling can promote knowledge of health risk situations.

Chirico et al. \cite{chirico2021designing} conducted a study showing statistical data on plastic consumption using VR. They compared three types of visualization: textual or numerical, concrete in the form of infographics, and mixed. 
In their study, they evaluated emotions, the sense of presence, and attitudes toward the use of plastic. Results indicated that the concrete and mixed formats were more effective than the numerical format, with similar results.

The use of infographics in AR has recently increased in visualization research. 
Yantong et al. \cite{Yantong_2020}, for instance, used AR to visualize the Beijing 2022 Winter Olympics infographics. 
Dehghani et al. \cite{Dehghani_2021} presented an AR-based infographics app for enhancing knowledge in biology classes. 
Chen et al. \cite{Chen2020, Chen2020These} proposed a set of tools to help non-expert users create self-authoring storytelling using infographics in AR.
Immersive infographics in VR also can be enjoyable. Romat et al. \cite{Romat_2020} report a study that assessed how users enjoy drawing pictographs in 2D and immersive in VR.

\subsection{\scalebox{.93}[1.0]{The Case for Immersive Concrete Scales Infographics}}
\label{sec:rel:imm_conc_info}

As concrete scales can help users understand complex quantities, immersive concrete scales can also increase the perception of large quantities. ``Visceral experiences'' in immersive environments can help users better understand quantities and scales  \cite{Lee_2020}. 
Lee et al. argue that \textit{data visceralization} offers us the possibility of ``being there'' and taking our conclusion about how immense famous monuments are, for example. 
In the same perspective, the literature has discussed VR applications for a deeper understanding of data regarding various social and environmental problems. Scurati et al. \cite{Scurati_2020}, for example, highlighted the potential of VR to provide motivational and informative experiences.

We believe the concepts of \textit{concrete scales} and \textit{data visceralization} are directly applicable to immersive concrete scales infographics, leveraging the user's physical environment as a reference of scale. Prior work has already successfully employed mobile AR as a tool to improve volume understanding in the context of food portion estimation \cite{stutz2014food,rollo2017servar}.

Concrete scales infographics in VR have the key advantage of being unconstrained by the reader's physical space. In this approach, it is possible to demonstrate, for example, the scale of buildings, even if the user is located in an indoor environment. Virtual objects can be added as context for the information, and the user is entirely focused on the information presented. 
On the other hand, AR infographics have the advantage of leveraging the reader's familiar environment and real objects as a reference for the information. They may also be considered more convenient by readers, as they do not require isolation from the physical environment. An additional advantage, particularly within the short-term future, is that AR infographics can be simulated through camera applications for mobile devices such as smartphones and tablets, currently much more widely available than headsets and already associated with news reading by a large number of people \cite{pew2012tablet, pew2021news}. 

The current uses of concrete scales infographics in the literature and in the news media (as discussed in Section \ref{sec:intro}) suggest two dimensions of particular relevance in their design space: the adoption of AR or VR and the adoption of distorted or real scales. While the opportunity to display the information on its real scale is the biggest motivation for an immersive perspective, in some cases, the amount of data will inevitably be too large to represent in most environments. While, in this case, the information could perhaps be visualized in AR by asking the user to move outdoors or to a window, another option would be to present a miniaturized concrete scales infographic. In this scenario, a miniature view could also include 3D virtual objects at the same scale for size reference, similar to a conventional concrete scales infographic. Such an example was demonstrated by Assor et al. \cite{assor2023waste} by rendering a virtual avatar and a virtual car next to a large number of trash bags, all in miniature sizes. The same scale distortion approach could be used to visualize microscopic particles such as viruses. The contribution of miniature or enlarged AR concrete scales models, in comparison to conventional infographics and real-scale ones, remains to be evaluated and is beyond the scope of this work.

Another interesting example of immersive AR concrete scales infographics was recently demonstrated also by Assor et al. \cite{Assor2024_ACM_JCSS}, with the goal of presenting waste accumulation. They argued that using AR to show such information would have the potential to help people understand the amount of waste they produce. They illustrated this idea through AR prototypes representing a series of examples, such as a week's worth of waste produced by a restaurant in the form of garbage bags; the amount of water used in a toilet, represented by one-liter bottles; the amount of mining waste generated to manufacture smartphones; and the amount of plastic cups accumulated over time. Assor et al. developed the waste visualization scenario in three formats: text, 3D image on a screen, and AR, similar to our study. They also evaluated users' emotional feelings over the formats through the PANAS questionnaire. While our work is closely related to theirs, our main focus is specifically on quantifying volume understanding under different conditions. In this context, our study also extends beyond waste visualizations in order to investigate volume understanding for different units of information and for different information scales. As a result, we understand that the two studies are largely complementary to each other.

\section{Experiment Design}
\label{sec:experiment}

\begin{figure*}[t]
\centering
     \subfigure[\emph{Trash \trash}]{\label{fig:N1}\includegraphics[width=55mm]{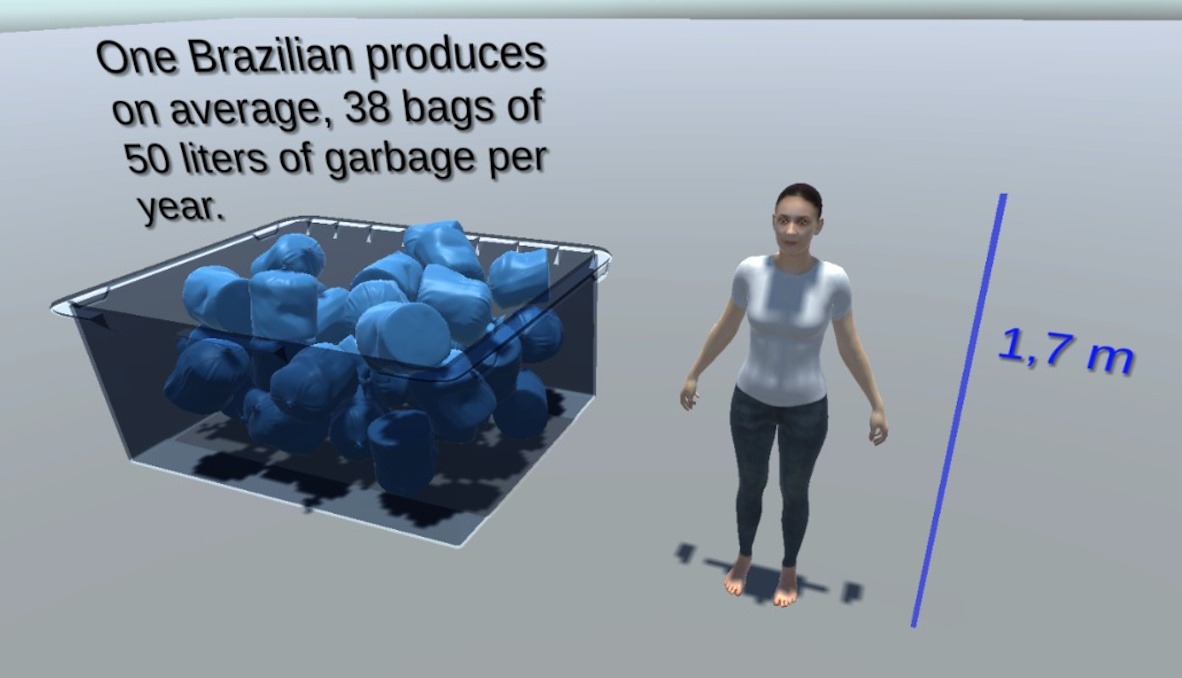}}
     \hfill
     \subfigure[\emph{Water \water}]{\label{fig:N2}\includegraphics[width=55mm]{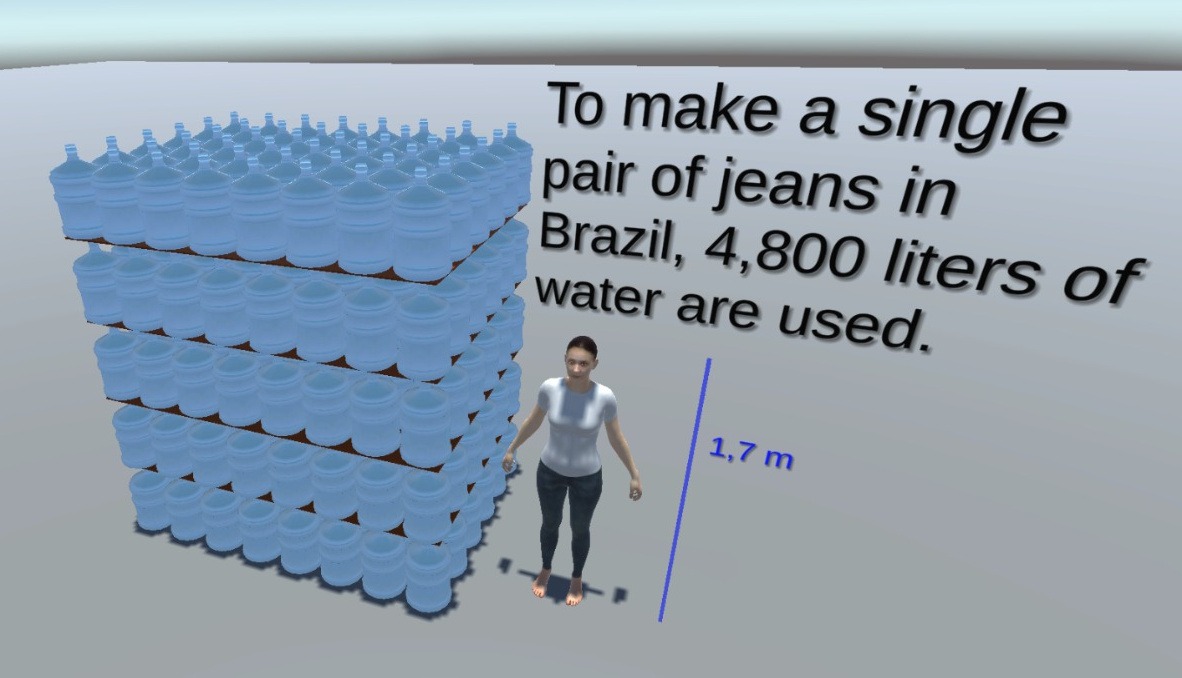}}
     \hfill
     \subfigure[\emph{Money \money}]{\label{fig:N3}\includegraphics[width=55mm]{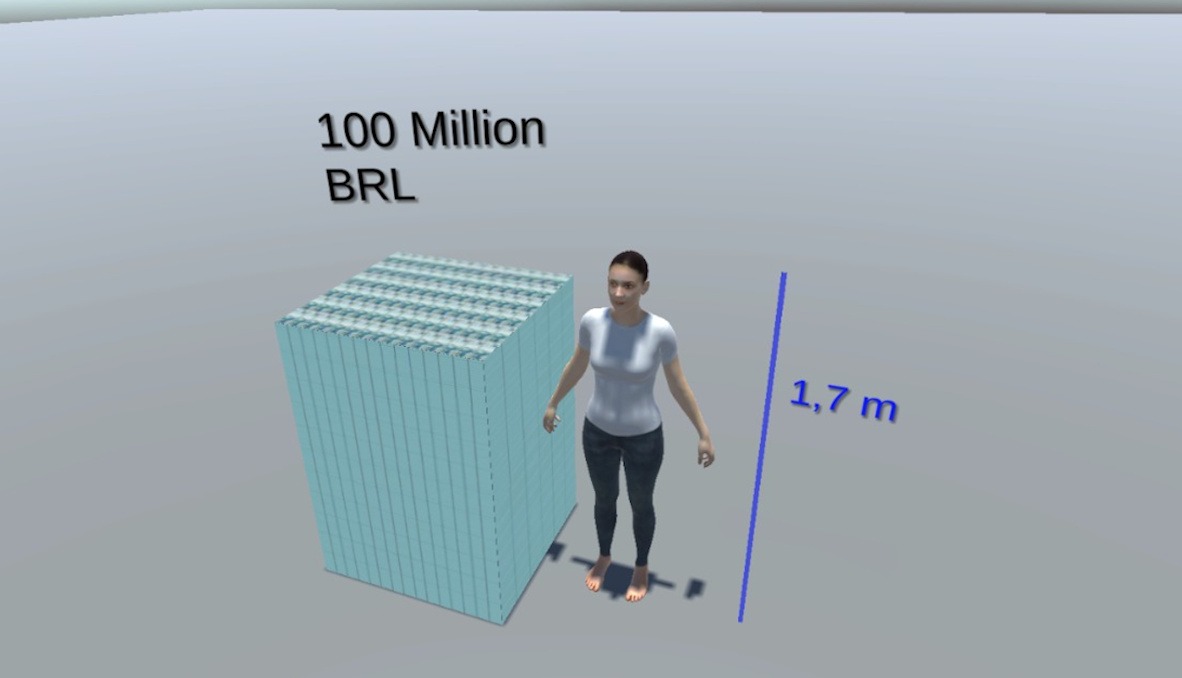}}
    \caption{The static infographics we generated for our three selected pieces of news employ familiar objects (50-liter trash bags, 20-liter water gallons, and R\$100 banknotes) to convey the information size. Following common practices in journalistic infographics, they include a 1.7m human representation to further contextualize the information size.}
 \label{fig:Prototypes}
\end{figure*}

\begin{figure*}[t]
\centering
    \subfigure[\emph{Trash \trash}]{\label{fig:N1_AR}\includegraphics[width=55mm]{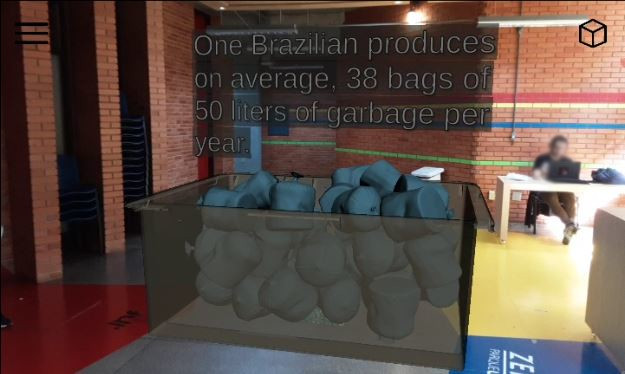}}
     \hfill
     \subfigure[\emph{Water \water}]{\label{fig:N2_AR}\includegraphics[width=55mm]{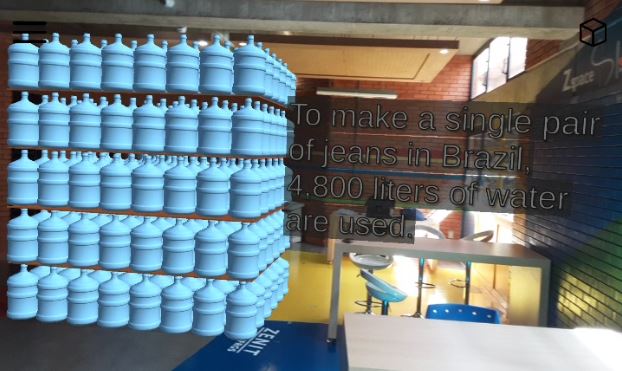}}
     \hfill
     \subfigure[\emph{Money \money}]{\label{fig:N3_AR}\includegraphics[width=55mm]{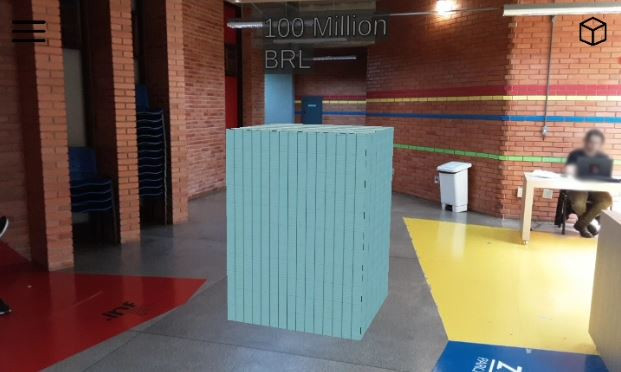}}
    \caption{The immersive concrete scales infographics we generated employ the same familiar objects used for the static infographics (\autoref{fig:Prototypes}). However, they are rendered in 1:1 scale in the user's real environment for exploration through our tablet-based mobile AR application. Note: while text contrast appears low in the screen captures, the participants considered it easily readable in our experiment.
        }
 \label{fig:AR_Prototypes}
\end{figure*}

While immersive concrete scales infographics can take multiple possible approaches, as discussed above, in this work, we are particularly interested in investigating their effects when users explore concrete scales metaphors in 1:1 scale through mobile devices in AR, allowing the use of the user's physical environment as a scale reference. To this end, we designed a set of three different study stimuli (Section \ref{sec:exp:stimuli}) based on different pieces of news corresponding to quantities of different units of measurement. We also designed a user study (Section \ref{sec:exp:study}) in which we asked 26 participants to indicate, through a mobile AR volume specification tool (Section \ref{sec:exp:tool}), the volumes they estimated after interacting with each stimulus in each of three possible representations (textual, image, and AR).

\subsection{Study Stimuli}
\label{sec:exp:stimuli}

To create a set of immersive infographics, we collected and adapted three examples of news articles presenting different units of quantities: kilograms of trash, liters of water, and amounts of money. We chose these three examples because the units involved can be difficult to estimate through numbers alone without any visual or analogy aid.
For each adapted piece of news, we generated textual summaries using analogies to familiar objects (shown in the boxes at the beginning of each section below) and developed an infographic prototype using the Unity3D game engine (\autoref{fig:Prototypes} and \autoref{fig:AR_Prototypes}).  
It is noteworthy that while the examples are presented here translated into English, in the study, they were shown in Brazilian Portuguese, the participants' native language.  

We developed an Android mobile application to support reading and visualizing the news. We used a marker to position the infographics in the camera field of view with the correct scale, as we needed to ensure that the virtual objects had an accurate concrete scale and not an approximated one. For that, we used the Vuforia software development kit. In this work, we could not use localization methods based on structure from motion using a monocular camera as these methods do not guarantee a correct scale \cite{Szeliski2022}. We used a Samsung S6 Lite tablet with a 10.4-inch screen to run the application.

\subsubsection{Waste production per capita (\textit{Trash} \trash)}
\label{trash_scenario}

\vspace{0.2cm}\begin{mdframed}
\textit{In 2019, 79.06 million tons of urban solid waste were generated in Brazil. Each Brazilian produced an average of 379.2 kg of garbage per year. This weight is equivalent to approximately 38 bags of 50 liters of garbage. 
Can you imagine how much space it would take to store 38 50-liter bags of garbage?} Adapted from \cite{PiauiFolha}.
\end{mdframed}

We often find news stories telling us a quantity in kilograms when, in fact, the objective is to inform a volume. Using a kilogram-based metaphor for garbage, for example, does not help the reader to imagine how much space that amount of garbage takes up. Thinking about the volume of something with just mass information is physically wrong since volume depends on the density of a material. Thus, if we use infographics to represent the volume, the reader can better understand the amount of garbage shown in the news reports. The garbage bags in Brazil are measured in liters, so we decided to use a 50-liter bag as a reference. This bag size is largely used at homes, and it can store an average of 10 kilograms of trash. While this chosen metaphor may not make sense for other countries, it could be easily adapted following the same logic but using different parameters of measures. 

Using this metaphor, we adapted our news example by adding the information that 379.2 kg corresponds to approximately 38 50-liter garbage bags. This way, we give the readers physical references to improve their comprehension. Using this example as a reference, we created a 3D model to depict this quantity of garbage (Figure \ref{fig:N1}). Garbage bags were stacked in a simplified manner inside a rectangular container to help participants understand and specify the volume in the study.

\subsubsection{Use of water in manufacturing (\textit{Water} \water)}
\label{water_scenario}

\vspace{0.2cm}\begin{mdframed}
\textit{Making a pair of jeans in Brazil consumes more water than a person in a month. According to the United Nations, a person uses about 3,000 liters of water per month, while manufacturing a single pair of jeans in Brazil uses about 4,800 liters, equivalent to approximately 240 gallons of 20 liters. Can you imagine how much space 240 gallons of 20 liters of water would occupy?} Adapted from \cite{PiauiFolha_moda}.
\end{mdframed}

Although liters are a unit of volume, if the reader does not know how much space a thousand liters take up, they will hardly understand how much water is needed to make a pair of jeans and how much water a person needs in a month. 
We thus rewrote this example by contextualizing that 5,200 liters correspond to 260 familiar 20-litter gallons of mineral water. These types of gallons are very popular in Brazil and should be easily recognized by the participants. Like in the garbage scenario, this metaphor could easily be replaced by other types of containers such as PET soda bottles.
We also slightly adapted the original information, changing the number of liters from 5,200 to 4,800 so that we could construct a prototype with a regular prism format, which we considered fairer in order to assess users in the experiment (Figure \ref{fig:N2}) in a way similar to the regular container used in the \textit{Trash} scenario.

As a consequence of the selected metaphors, in both \textit{Trash} and \textit{Water} scenarios, the infographics' total volumes are larger than the minimum possible volume occupied by the amount of water or trash mentioned in the news. This is due to the empty spaces around the objects' irregular shapes when stacked in a natural way. Our intention was to measure the understanding of the estimated volume of these concrete scales metaphors, and even in the textual analogies, participants were asked to imagine a stack of these kinds of objects in their estimations.

\subsubsection{Amount of money in a lottery prize (\textit{Money} \money)}
\label{money_scenario}

\vspace{0.2cm}\begin{mdframed}
\textit{A prize of R}\$ \textit{100 million from the \textit{Mega-Sena} lottery can yield a monthly income of R}\$ \textit{896 thousand for
the winner. If you invest the prize money in a simple fixed-income investment, a certificate of deposit that pays 100\% of the CDI index, with the current interest rate at 12.65\%, the prize will yield R}\$\textit{ 10.7 million in the first year. Can you imagine how much space 100 million reais in R}\$ \textit{100 banknotes would occupy?} Adapted from \cite{MegaSena}.
\end{mdframed}

\begin{figure*}[t]
\centering
     \subfigure[Text]{\label{fig:Step1}\includegraphics[width=52mm]{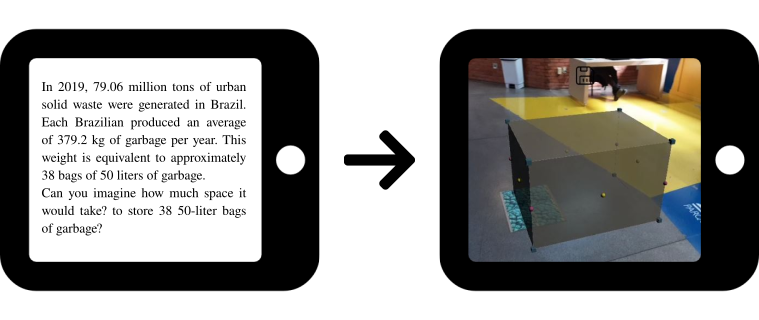}}
     \hfill
     \subfigure[Image]{\label{fig:Step2}\includegraphics[width=52mm]{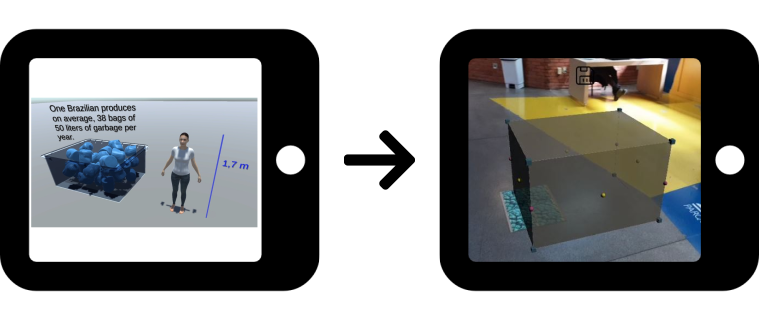}}
     \hfill
     \subfigure[Mobile AR]{\label{fig:Step3}\includegraphics[width=52mm]{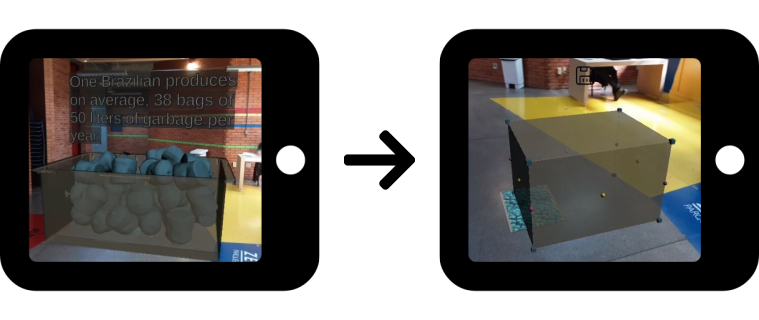}}
    \caption{As an example of the study workflow, a participant in Group A performs the tasks from the Trash scenario after the Training in the following order: (1) the user is presented with the Text format, then draws the prism in AR to estimate the volume reported in the text; (2) the user is presented with the static infographic as an Image and then draws the prism in AR to estimate the volume reported in the static infographic; (3) the user is presented with the immersive infographic in mobile AR, can walk around the infographic, and then draws the prism to estimate the volume reported in the AR infographic. \autoref{tab:tasks_order} details the counterbalanced condition orderings for all groups.}
 \label{fig:trash_steps}
\end{figure*}

It is common to find news articles referring to vast amounts of money, as illustrated by this one on Brazil's most famous lottery prize. It is intuitive to think about things one can buy with that money. Another common way of representing large amounts of money is to describe how much space it would take up if we stacked familiar banknotes. 
For this scenario we decided to use money in R\$100 banknotes because this metaphor is a classic example already discussed in prior concrete scales infographics work \cite{Chevalier_2013,Lee_2020}, a consequence of the fact that large quantities of money can be considered hard to conceive. We adapted this excerpt to ask participants about the space that 100 million Brazilian Reais (BRL) in R\$100 banknotes would occupy (Figure \ref{fig:N3}). The value of 100 million was also selected so that the resulting representation would fit our selected test environment. During development, we experimented with different examples and concluded that a larger amount, such as 1 billion, would be too large for users to view on a tablet on a real scale, compromising the immersion of the visualization.

\subsection{User Study Design}
\label{sec:exp:study}

We designed a within-subjects study to test our set of infographics (i.e., Trash \trash, Water \water, and Money \money). With them, we intended to measure to what extent a mobile AR visualization can help people understand volume quantities better than conventional approaches. 

Therefore, for each infographic, we provided three different manners of reading and visualizing the information: (i) a simple text describing the news with analogies to familiar objects (see the text box of each scenario at the beginning of each subsection in Section \ref{trash_scenario}), (ii) an image showing a 3D infographic about the news containing a human-sized reference, as shown \autoref{fig:Prototypes}, and (iii) an immersive concrete scale infographic in mobile AR showing the same infographic of the image but in the user's environment and supporting immersive exploration (\autoref{fig:AR_Prototypes}).

Since our goal is to evaluate each visualization approach under its expected usage mode in a real news reading scenario, in the immersive condition, users are not constrained to the same viewing angle predefined by the static image infographic. Instead, they are allowed to walk around to get a better understanding of the space occupied by each 3D model, as they hypothetically could also do in their homes.
However, we decided that the participants should all take the tests in the same place, a common environment often used by participants at the university so that everyone would have the same physical objects as a spatial reference.

\subsubsection{Hypotheses}
\label{sec:exp:hypos}

Recent works on the use of infographics in immersive settings have mainly assessed emotions \cite{chirico2021designing, assor2023waste, Assor2024_ACM_JCSS}. However, the main idea of using concrete scales infographics is to convey the exact or best possible estimated value of the data being shown to spectators through some metaphor they know well \cite{Chevalier_2013,Lee_2020}. 

In this work, our aim is to assess how much infographic visualization in mobile AR helps readers understand quantities. We opted to conduct our study to estimate only volumes for evaluating three visualization formats  (Text, Image, AR) in three scenarios based on news pieces (Trash, Water, Money) to maintain the experiment within a limited time duration. Since what matters is to use a metaphor based on common knowledge, volume estimation seemed to be a natural choice for representing the quantities mentioned in the news. Furthermore, we also measured participants' feelings using a specific metric for infographics. 
We defined three hypotheses for our study:

\begin{enumerate}

\item[H1] Interacting with a mobile AR infographic will lead to a smaller volume estimation error than an Image only.
\item[H2] Viewing an Image infographic will, in turn, lead to a smaller volume estimation error than textual analogies only.
\item[H3] Interacting with a mobile AR infographic will lead to a higher degree of \textit{concerned}, \textit{sadder}, and \textit{awestruck} reported feelings compared to both Text and Image.
\end{enumerate}

\textbf{H1} expresses the main objective of this study and corresponds to our speculation that immersive concrete scales infographics should improve the understanding of volume information by allowing the use of the physical environment as a reference. \textbf{H2} reinforces the expressivity power of infographics in general, positing that both immersive and non-immersive ones will contribute to the user's understanding. In addition to the quantitative measures, we will also verify the participants' emotional reactions through \textbf{H3}, and we believe they will feel more impacted by the immersive concrete scales infographics than by the other two conditions.

\begin{figure}[t]
\centering
    \includegraphics[width=\linewidth]{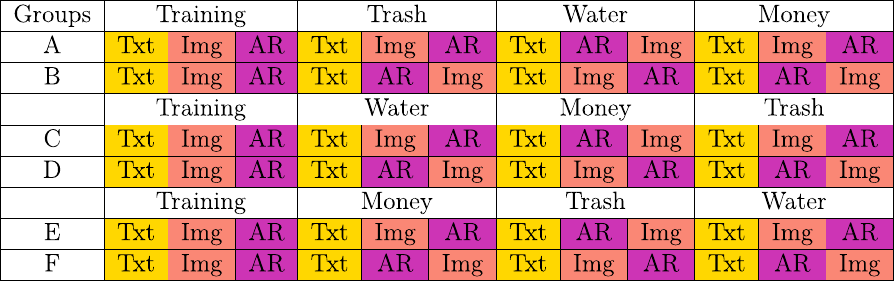}
    \caption{We separated the participants into six groups, each one doing the tasks in a different order to balance our user study. \color{black}{We deliberately always showed the Text format first in order to capture the participant's original assessment prior to receiving any visual indications through infographics, thus serving mostly as a baseline result for the two visual conditions.}}
 \label{tab:tasks_order}
\end{figure}

\subsubsection{Procedure}
\label{sec:exp:proc}

\begingroup

We recruited 26 graduate and undergraduate student volunteers from our University (17 male, 8 female, and 1 gender not informed, mean age 25.15, SD 6.5). 
\setlength{\columnsep}{5pt}%
\setlength{\intextsep}{5pt}%
\begin{wrapfigure}{r}{0.13\textwidth}
    \centering
    \includegraphics[width=0.13\textwidth]{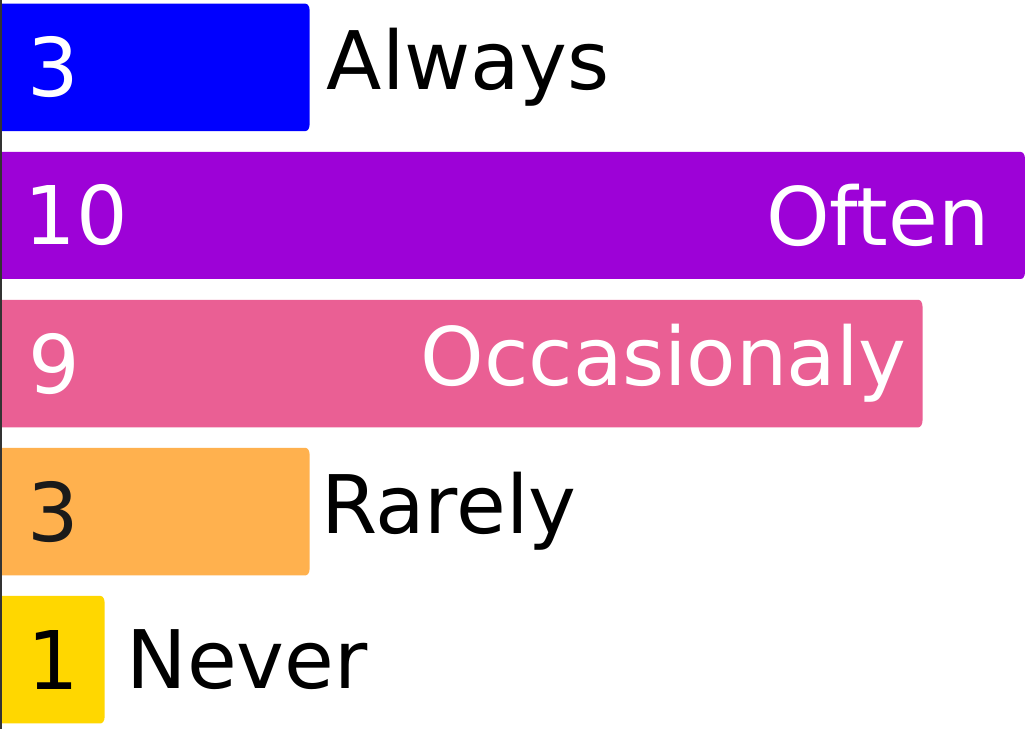}
\end{wrapfigure}
10 participants reported never having used AR before, while 14 had already used mobile AR, and 2 had already used it through other devices such as headsets.
Moreover, 22 participants reported being used to reading the news on their phones at least occasionally (see inline figure). 

\endgroup

The three pieces of news were presented to participants in counterbalanced order using the Latin square, and so were the visualization formats except for Text, which the participants always viewed first. Our reasoning for that was that viewing any infographic (either Image or AR) before the textual version would inherently compromise the user's original assessment, and results would not be accurate for the text modality.  It is important to emphasize that the textual condition should not provide much information for the subsequent modes, as the participants are not shown any confirmation of how correct their estimation was. Using this approach, we could also better measure the surprise effect of text modality versus the other modalities. Moreover, it should be noted that our main focus is on comparing immersive and non-immersive infographics, and thus, we view the Text condition mostly as a baseline to contextualize those results. 
\color{black}
In total, we had six possible order combinations groups as shown in \autoref{tab:tasks_order}. Also, as previously mentioned, all the participants performed the tasks in the same controlled space, which included some desks, chairs, trash cans, and poufs as references.

After viewing the news in each format (Text, Image, and AR), we asked the participants to interactively adjust a rectangular prism in mobile AR to represent the volume reported in the news (\autoref{fig:trash_steps}). With the estimated volume provided by the users, we expect to measure how well each visualization modality helps participants understand the actual volumes described in the news.

The training was executed in the same order for all groups, and the tasks were performed in the different orders detailed by \autoref{tab:tasks_order} (\autoref{fig:trash_steps} illustrates the order for Group A as an example). For the training for the Text format, we asked participants to imagine a chair and then draw a prism with the size of the chair that they had imagined (Figure \ref{fig:Step1}). Then, we showed them an image of a chair next to a person with their height indicated, as in the static infographics we designed for the actual study. Then, they were asked to draw a prism to replicate the same volume indicated in the image (Figure \ref{fig:Step2}). For the last training step, we showed a chair in mobile AR and then asked participants to draw a prism that replicated its size (Figure \ref{fig:Step3}).

In addition to the volume estimation tasks, we asked participants how they felt when visualizing and reading each news piece in the three visualization modalities to gather their feelings and evaluate whether they changed depending on the visualizations. For this, we used the \textit{Smile or Scowl?} infographic taxonomy developed by Lan et al. \cite{Lan_2021}, who categorized feelings into twelve moods (six smiles and six scowls) as a result of two subsequent user studies where participants reacted to a set of 976 infographics on different topics.

\begin{figure}[t]
\centering
    \includegraphics[width=\linewidth]{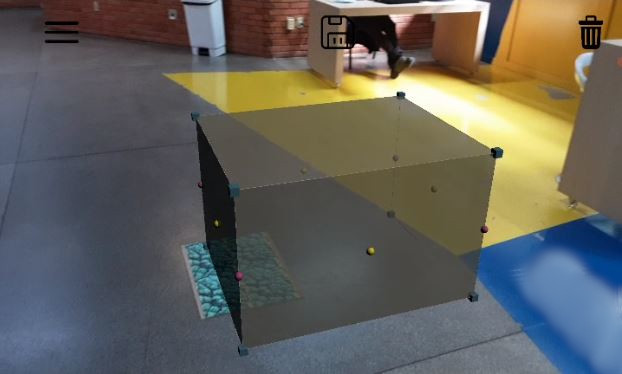}
    \caption{Our AR volume specification tool allowed users to translate, rotate, and scale a semitransparent rectangular prism to indicate their estimations for the volume informed in the news. Real-world objects available in the environment could serve as size references.}
 \label{fig:cube}
\end{figure}

\subsection{Volume Specification Tool}
\label{sec:exp:tool}

As introduced in Section \ref{sec:exp:proc}, after reading and viewing the news piece in each of the three modalities, Text, Image, and AR (the order of the last two being always counterbalanced), we asked participants to draw, using a mobile AR tool, a rectangular prism to represent the volume discussed on the news. We used the MRTK interaction toolkit \cite{MRTK} for Unity3D to implement this tool, enabling efficient prism manipulation through widgets. 
\autoref{fig:cube} shows an example of a rectangular volume prism. Users may change the size of the prism by dragging its corners (blue cube widgets). With the pink sphere widgets in the vertical edges, participants can rotate it around the \textit{y}-axis. Finally, with the yellow sphere widgets in the middle of each prism face, it is possible to move the prism along the floor plane. With these three features, participants can adjust the size of their prism, rotate and translate it to bring it closer to real objects in the environment, and use them as a spatial reference to provide reasonable volume estimations. The system automatically records the final selected volume.

\begin{figure*}[t]
\centering
    \includegraphics[width=\textwidth]{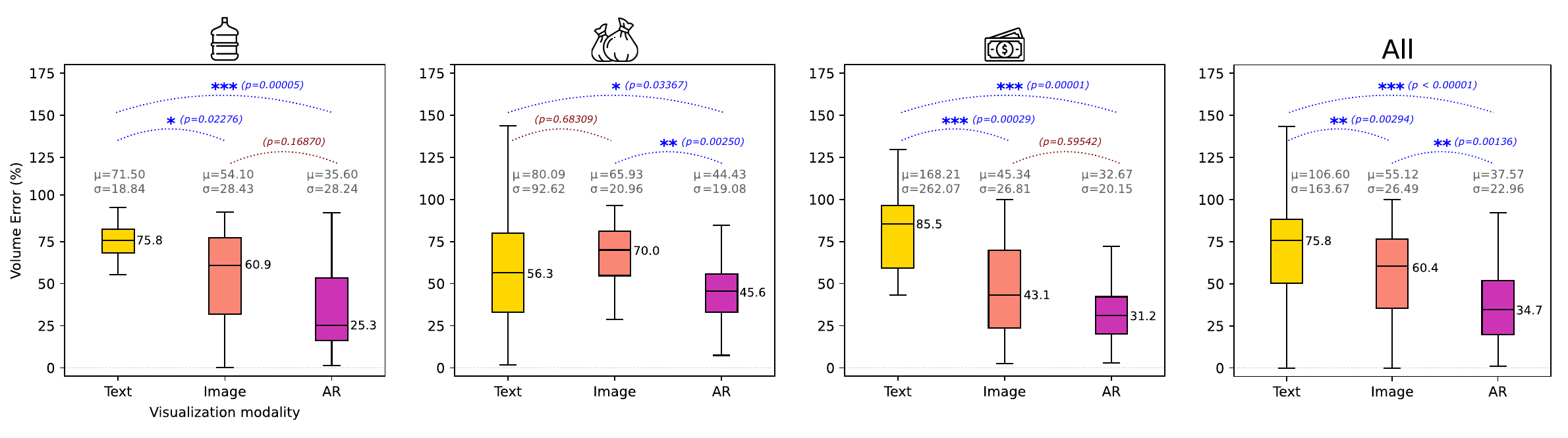}
    \caption{Results of the error (percentage) in volume estimation for each piece of news. Considering the combination of all scenarios, AR led to a significantly smaller estimation error compared to both other modalities, and the same was also observed in the Trash \trash~scenario specifically. Both types of infographics led to significantly smaller estimation errors compared to textual analogies alone ($\mu =$ average and $\sigma =$ standard deviation).}
 \label{fig:volume_error_by_news}
\end{figure*}

\begin{figure*}[t]
\centering
    \includegraphics[width=0.75\textwidth]{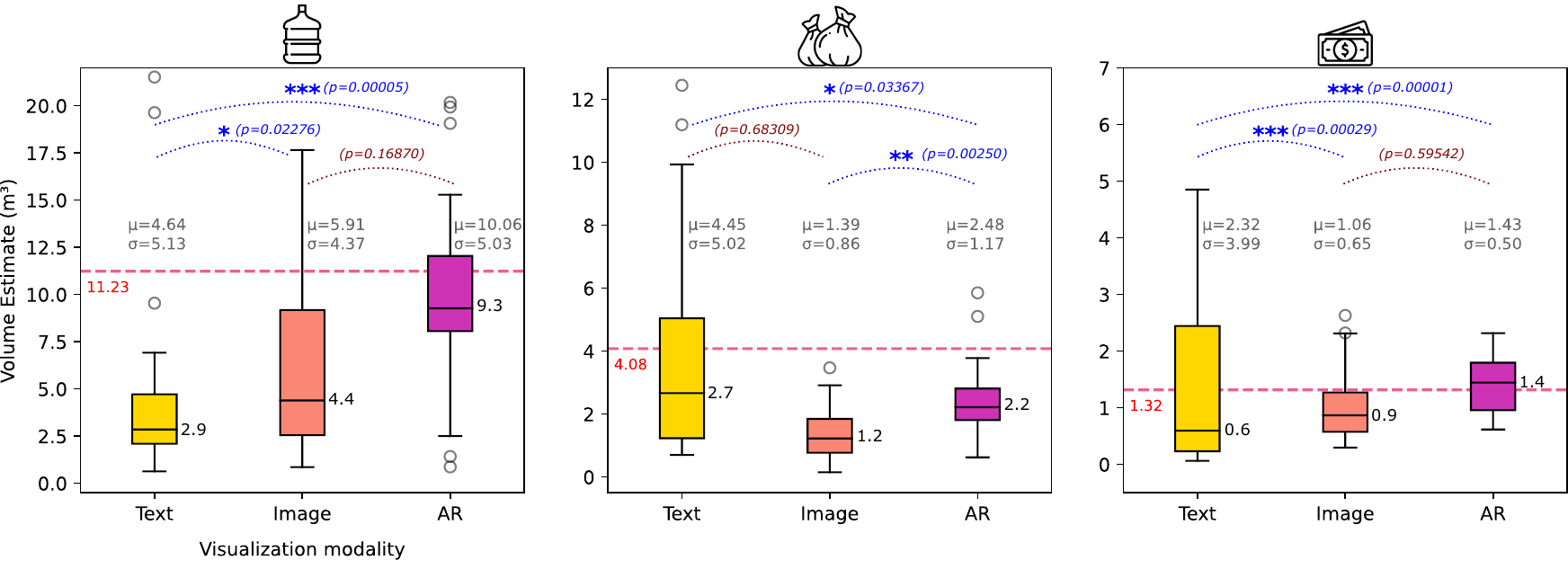}
    \caption{Results of volume estimation in $m^3$ for each news piece: \textit{Water}--left, \textit{Trash}--center, and \textit{Money}--right, presented from the largest to the smallest. The pink lines indicate the correct volume of each 3D infographic representation. Note that each plot is on a different \textit{y}-scale.}
 \label{fig:volume_estimate_by_news}
\end{figure*}

\section{Results and Findings}
\label{sec:results}

In this section, we report the results of our comparative user study regarding volume understanding, infographic-induced feelings, and users' perceptions. For increased readability, we structured this section around our most relevant findings. 
For the statistical comparisons, as our samples were not parametric, we used the Friedman test with the Wilcoxon-Nemenyi-McDonald-Thompson post-hoc test \cite{hollander1999nonparametric}. The significance is indicated as follows:
\begin{center}
    (*) for $p < 0.05$, (**) $p < 0.01$ and (***) $p < 0.001$.
\end{center}

\subsection{AR infographics consistently led to smaller volume estimation errors (H1 \cmark)}

\autoref{fig:volume_error_by_news} shows the percentage of volume estimation error for each one of the news examples and visualization modalities, as well as an aggregation of all study trials across examples. Meanwhile, \autoref{fig:volume_estimate_by_news} shows the distribution of the absolute volume estimates for each news item with their respective real volume values indicated by a pink line. 
We were particularly interested in observing how AR would compare to Image, i.e., to what extent AR infographics would be more helpful for volume understanding than conventional infographics.

\begin{figure*}[t]
\centering
    \includegraphics[width=\linewidth]{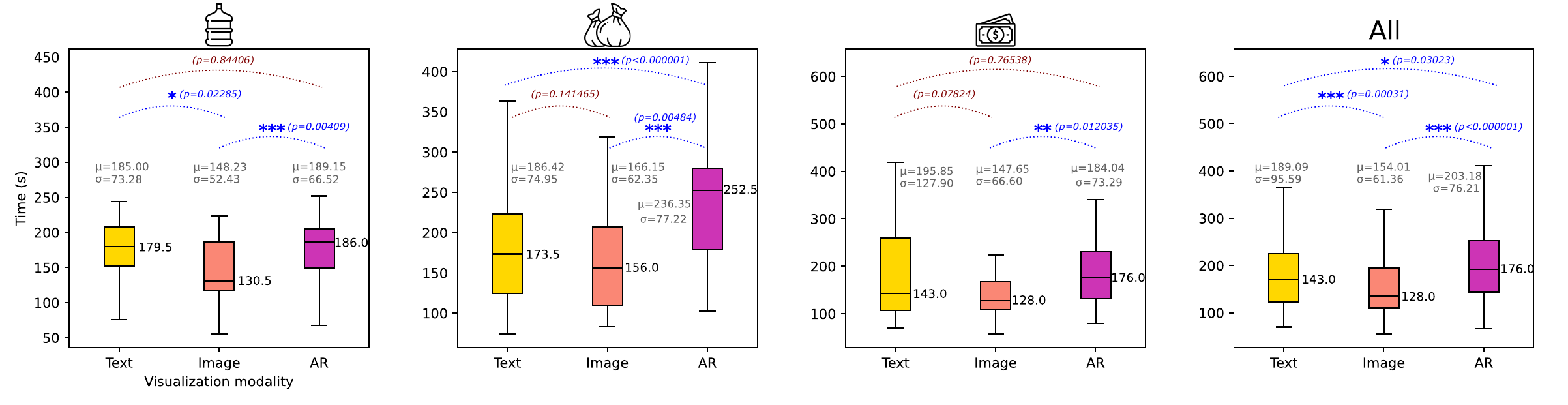}
    \caption{{Total task completion times in seconds for each piece of news. Task completion comprises the news consumption, the content understanding, the reasoning about the quantities communicated, the interaction with the text/infographic, and the volume estimation. When analyzing all news scenarios combined, Text led to longer times than Image by requiring cognitively estimating volumes without any visual cues, while AR led to the longest times due to the need for spatial interaction.}}
 \label{fig:time_by_news}
\end{figure*}

\begin{figure}[t]
    \centering
    \includegraphics[width=\linewidth]{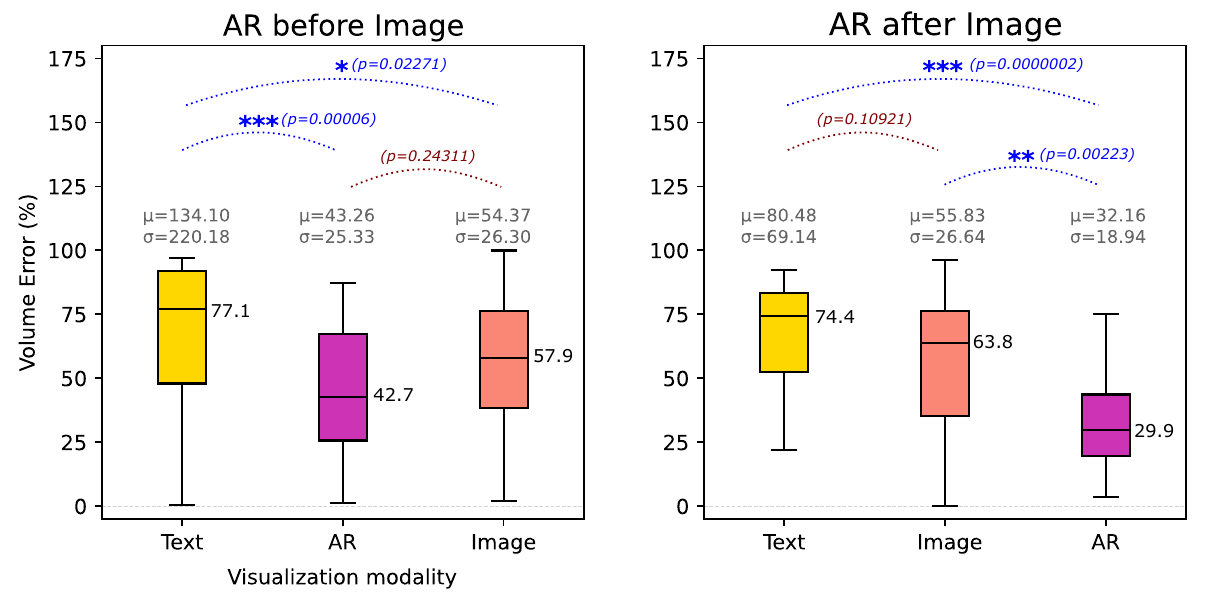}
    \caption{Results for all tasks combined divided into two groups (\textit{AR before Image} and \textit{AR after Image}) yielded different patterns, suggesting a possible learning effect due to order. While AR always showed the smallest error volume estimate, AR and Image were significantly different only when Image was experienced first. This suggests that seeing the AR infographic first was apparently more helpful than the opposite way.}
 \label{fig:order_effects}
\end{figure}

Grouping all tasks together, as shown in \autoref{fig:volume_error_by_news}--right, the percentage of error is significantly different across all formats. Users were more accurate in estimating volumes for news presented using Image than Text (**); the same holds for AR when compared to Text (***) and to Image (**). 
While Image resulted in an average error of 55.12\%, AR led to an average estimation error of 37.57\%, i.e., a relative reduction of 31.8\%. 
These findings confirm H1, indicating that AR outperforms Text and Image for volume understanding. 

We also analyzed the three different news scenarios individually. 
In the \textit{Trash} scenario (\autoref{fig:volume_error_by_news} \trash), our study was also able to identify a significant improvement in estimation accuracy when using AR in comparison to Image (**). In this scenario, Image led to an average volume estimation error of 65.93\%, while AR resulted in 44.43\%, a reduction of 32.6\%. In the \textit{Water} and \textit{Money} scenarios, we could not say AR was better than Image, as we could not find a statistically significant difference between them. However, the volume estimation of the \textit{Money}--AR pair was the nearest to the correct volume, as shown \autoref{fig:volume_estimate_by_news}.
Analyzing both the \textit{Water} and \textit{Money} scenarios, Image led to an average volume estimation error of 54.10\% and 45.34\%, respectively, while AR had 35.60\% and 32.67\%. Therefore, for the \textit{Water} and \textit{Money} scenarios, we had a volume estimation error reduction of 18.4\% and 12.67\%, respectively.

\subsection{Both infographics types led to a better comprehension compared to textual analogies alone (H2 \cmark)}

When aggregating data from all case studies, both Image and AR present significantly lower error rates compared to Text. While Text led to an average estimation error of 106.6\%, Image resulted in an average error of 55.12\% (**, a relative reduction of 48.3\%) and AR, 37.57\% (***, a reduction of 64.8\%). \autoref{fig:volume_error_by_news} further reports on the mean and standard deviation statistics. 
This result confirms the contribution of both immersive and conventional infographics for volume understanding, thus confirming hypothesis H2. This is not entirely surprising given the long tradition of the use of infographics to complement textual information.

Considering the analysis of the scenarios individually, Image notably led to significantly smaller volume estimation errors compared to textual analogies alone in all scenarios except for \textit{Trash}. AR led to significantly smaller errors compared to Text for all scenarios (\autoref{fig:volume_error_by_news}).

\begin{figure*}[t]
\centering
    \includegraphics[width=\textwidth]{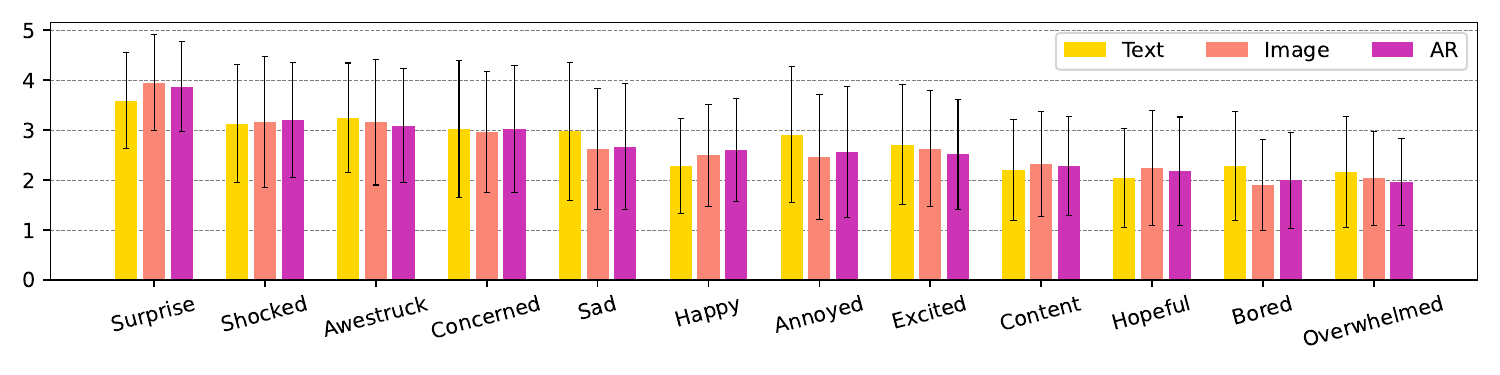}
    \vspace*{-8mm}
    \caption{Distribution of users' feelings obtained through the \textit{Smile or Scowl?} questionnaire related to the three formats for all three scenarios.
    Contradicting our hypothesis H3, no major differences were observed between conditions in terms of infographic-induced feelings.}
 \label{fig:feelings_by_news}
\end{figure*}

\begin{figure}[t]
\centering
    \includegraphics[width=\linewidth]{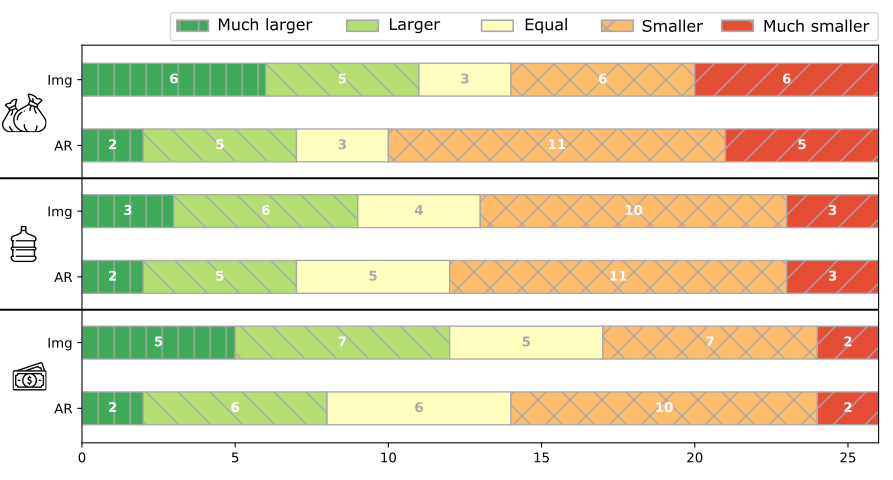}
    \caption{\color{black}\textit{``How big or small was the
amount you imagined when reading the news before visualizing it through this infographic?''} The green bars on the left (\textit{much larger/larger}) indicate participants who reported overestimating the size, while red ones (right side, \textit{smaller/much smaller}) indicate underestimations. The yellow bars (center, \textit{equal}) indicate that the participant reported imagining the same size as the infographics later showed. In general, participants reported tending to more often underestimate than overestimate the size of the information received before seeing the infographics.}
 \label{fig:question4}
\end{figure}

\subsection{Possible order effect corroborates the larger contribution of AR infographics to volume understanding}

We also verified a possible learning effect due to the order in which Image and AR visualization modalities were presented to the users as part of the study counterbalancing. 
\autoref{fig:order_effects}-left (\textit{AR before Image}) and \autoref{fig:order_effects}-right (\textit{AR after Image}) show the results of the analysis performed with the users divided into these two groups. We observe that when the news was presented in the \textit{AR after Image} order (\autoref{fig:order_effects}--right), we cannot say that the users were more accurate with the Image than with the Text visualization, but the results with AR are significantly better than both with Text (***) and Image (**). On the other hand, \autoref{fig:order_effects}--left shows that the Image visualization is more accurate than the Text (*), and AR is also better than Text (***) when the AR visualization is presented before the Image. However, this time, we cannot affirm that AR is significantly better than the Image visualization for volume estimation.

Analyzing these results, the Image visualization is only better than the Text when presented after the AR modality. Likewise, Image is not significantly worse than AR when the latter is presented before the former (unlike the opposite case). This suggests a potential learning effect, with AR increasing the performance of the Image visualization as the user can memorize the AR visualization and use it to perform the estimation based on the Image modality.

\subsection{AR infographics led to the longest task completion times, and static image ones to the shortest}

\autoref{fig:time_by_news} shows the average total task completion times for each scenario and condition, as well as for all scenarios combined. This includes the time from the moment the participant started interacting with the infographic or text to the moment they submitted their prism volume estimate. We opted not to decompose the task completion times into ``interaction time'' and ``estimation time'' as each participant organized the task execution differently, sometimes starting the estimation cognitive process while still viewing the text or infographic. We further argue that the effort to construct the prism is constant, and any differences between conditions should be attributable to the affordances of the different visualization modalities. 

While the data from \autoref{fig:time_by_news} shows that the shortest task completion is achieved using static 2D infographics in all scenarios, this is not an indication that it is also the best visualization modality. By combining the times measured and our observation of the experiment sessions, we noticed that the Text modality takes longer due to the time the users took reasoning about the volumes without the help of any visual cue. Conversely, the long time spent in the AR modality was due to the exploration of the data in the physical space. The users move around, exploring the 3D infographics from different perspectives and looking for cues in the real environment that help them to draw the prism. While, in the Text modality, the long time spent may be seen as negative (it did not help to better estimate the volume), in the AR modality, it is positive. The user engaged with the data, which led to a more precise volume estimation (\autoref{fig:volume_error_by_news}). 

Indeed, we should say that shortening the time spent by users consuming news is not a goal. Usually, news media prioritize engaging readers by providing them with the best user experience possible, which might be obtained with AR infographics.

\subsection{Readers tended to underestimate volume sizes before seeing them depicted visually}

\autoref{fig:question4} depicts how participants' expectations before seeing each kind of visualization (i.e., Image and AR) compared to the information in the text only.
The possible responses were: much larger, larger, equal, smaller, much smaller. While shades of green (much larger, larger) indicate that participants subjectively acknowledged having initially overestimated the sizes, shades of red (smaller, much smaller) correspond to the opposite.

Analyzing the results for the \textit{Trash} scenario, we can see that, before viewing it in AR, 16 participants thought the volume was smaller or much smaller than it actually was, whereas, in the Image visualization, the feedback was more balanced, and only 12 thought the volume was smaller or much smaller. However, when we analyze \autoref{fig:volume_estimate_by_news}, we notice that, in reality, in the Image modality, the participants estimated the volume as smaller than in the AR modality---$\mu = 1.39$ and $\mu = 2.48$, respectively.
In \autoref{fig:question4}, we can see that the task with the most votes for \textit{equal} was \textit{Money}--AR, and checking in \autoref{fig:volume_estimate_by_news}, the \textit{Money}--AR task had in fact the best volume estimation. The real value was $1.32~m^3$, and the estimated average was $\mu = 1.43~m^3$. In this case, the subjective and quantitative perceptions were consistent.

\subsection{No significant differences were observed in terms of infographic-induced feelings (H3 \xmark)} 

\autoref{fig:feelings_by_news} shows the \textit{Smile or Scowl} 5-point Likert scale questionnaire results aggregated for all news scenarios. According to these results, the means obtained for the three conditions were always similar, and therefore our hypothesis H3 cannot be confirmed at this point. 

Still, we noticed a minor non-significant trend: while the participants looked at the news, in part contrarily to our expectations, they felt the most \textit{sad}, \textit{bored}, and \textit{annoyed} in the Text mode. On the other hand, Image and AR tended to induce more \textit{surprise} reactions than Text. Such potential differences are inconclusive and should be better investigated in future work.

\subsection{Participants' feedback suggests interest in the use of AR infographics to support news pieces}

When we asked the participants what they thought about viewing the news in AR, they were generally very positive. For example, P6 stated that it was \textit{``Very interesting, the opportunity to view news is curious and attractive for different audiences, a new approach, and vision of the same.''}, while P10 affirmed \textit{``Interesting to have a more realistic view of the news.''}. We asked if they thought that the immersive visualization of the infographics helped with their perception of the volume, and one of the participants said \textit{``It makes much sense for news that involves the environment. It is quite difficult to visualize quantities like trash in normal news.''} [P7]. 

Participants also offered some qualitative feedback concerning the study interface. When asked about the volume estimation tool, they said that \textit{``It would be better to have a button to change the size of each face of the prism individually.''} [P25], \textit{``It would be nice to be able to lock the prism manipulation axis.''} [P8], and \textit{``Apart from the difficulty of handling the prism, it is an exciting tool for understanding quantities.''} [P17].

\subsection{Participants demonstrated similar strategic behaviors during the experiments}

When estimating the volume using the AR tool, we noticed two main behaviors of the participants. When estimating the volume after reading the news in the Text condition, some of them imagined and spoke out aloud about the suggested unit and mentally built a 2D matrix on the floor. Only later, when they viewed the infographics, they realized that they could stack the objects, and from then on, they started to think about 3D matrices. Another interesting behavior was that some users, while drawing the prism and moving when the prism rotated, also lowered their bodies to better see the height and check whether it was accurate by comparing it with the objects in the environment (e.g., tables and chairs).

\section{Discussion: Contributions, Limitations, and Perspectives for Future Studies}
\label{sec:discussion}

Our findings allowed us to identify clear benefits of immersive concrete scales infographics for volume understanding against conventional infographics for both the \textit{Trash} scenario and across all scenarios. Moreover, we could confirm that 
concrete scales infographics, in general, always significantly outperformed textual analogies alone. These findings are a relevant contribution to the literature on concrete scales infographics and immersive concrete scales infographics and also support the current trend of the growing use of AR visualizations in the news media.

Considering the fact that we have investigated three different examples of news articles conveying information of varying scales, we believe these findings are also likely to be confirmed for other information domains by employing comparable metaphors. Regardless, it should be noted that these findings constitute an initial investigation of immersive concrete scales infographics, and further research is needed to better understand the effect of associated factors. In particular, we can identify the following components as potentially linked to the findings obtained in our study. 

\textbf{Study design.} In our user study, we asked participants to estimate the information volume in the same environment where they viewed the immersive infographics. One could claim that this could introduce bias in favor of this condition because it would allow users to memorize relationships between the information and the space. We argue, however, that this is the key rationale of immersive concrete scales infographics by design: it allows people to use their environments and familiar physical objects as references to understand the information they are receiving. Regardless, future studies should investigate to what extent the gained comprehension is retained once the user moves to a different environment. 

\textbf{Choice of condition counterbalancing.} In the design of our experiment, we deliberately decided not to fully counterbalance our conditions because we considered that assessing textual analogies alone would only make sense before the user had seen a visual depiction of that same information. As a result, we always presented text analogies first, and this result served mostly as a baseline for the other two. It is important to emphasize that the textual condition provided limited information for the subsequent modes, as the participants were not shown any confirmation of how correct their estimation was. Further, the added gain in system experience should be minimal given that a training session had already been performed in the beginning of the experiment. Regardless, we acknowledge that this decision should be kept in mind when assessing our second hypothesis. Exploring full counterbalancing, or any other counterbalancing approach, could be an interesting direction for future work.

\textbf{Choice of study device.} In our study, we opted to use a tablet device for all conditions. According to a 2012 study from Pew Research, tablets are more commonly used for in-depth reading of news articles than smartphones \cite{pew2012tablet}, and they also offer better support for AR due to the larger display. However, future studies could also evaluate if similar findings extend to smartphone AR, as these devices are also widely used for information consumption. Similarly, they could also assess the use of AR head-mounted displays, which are expected to become more common in the future.

\textbf{Technical limitations.} While our tablet-based AR prototype worked well, as confirmed by our findings favoring AR, technical limitations in the environment and marker tracking sometimes led to minor glitches that may have caused user frustration and added noise to the subjectively reported infographic-induced feelings, for example. Future studies can investigate the effect of such limitations. Moreover, newer mobile devices, such as those including LiDAR scanners, are likely to improve the AR experience, potentially leading to even better results for AR infographics. Devices with more processing power can also permit more photo-realistic infographics with a similar expected impact. 

\textbf{Choice of volume specification apparatus.} In our study, we provided participants with an AR volume specification tool to indicate their estimations for the volume of information reported in the news. We considered this approach ideal as it allowed users to fully control their volume specification, visualize the resulting estimation, and compare it to their expectations. Given that users had the opportunity to train the use of the tool beforehand, we do not anticipate a significant bias being introduced by this tool in favor of any condition. Still, future studies could experiment with alternative measuring systems. 

\textbf{Choice of data analogies.} When designing an immersive or conventional concrete scales infographic, a relevant design choice is determining the most appropriate metaphor. For example, small water bottles, larger water gallons, or simply a cubic water tank could represent liters of water. Other quantities, such as garbage, are particularly difficult to represent, as the information is often conveyed in terms of mass instead of volume, and a visual representation must select an approximate density, which may differ from the reader's expectation. In all our examples, we iteratively worked to select the metaphors we deemed most adequate for representing the scale of the data while remaining widely familiar to the participants. Our observation from the study's results is that this was achieved. However, it is conceivable that alternative analogies could lead to different performances, and future studies should investigate this. 

\textbf{Design of the static infographics.} For the Image condition, we sought to reproduce a style commonly found in media infographics, representing quantities in terms of familiar units and contrasting them to a referent of known size. We organized these objects in a 3D space and offered a 3D perspective to clearly indicate all dimensions since we considered that a purely 2D infographic, while still realistic, would make it even harder for the participants to comprehend the information in terms of volume. Investigating alternative designs (for example, with different perspectives or different referents) for the Image condition could also be the object of future studies.

\textbf{Design of the immersive concrete scales infographics.} Similarly, numerous possible designs could be conceived as immersive concrete scales infographics, adopting different metaphors, 3D models, or spatial organizations. Future studies should investigate the most effective designs for such representations. As discussed in Section \ref{sec:rel:imm_conc_info}, immersive concrete scales infographics can also assume other forms, such as miniature-scale comparatives, which also warrant comparisons to full-scale ones.

\textbf{Choice of study environment.} Our user study was conducted in a university hall environment, a relatively large space that could accommodate representations of all scenarios. We selected this location as it represented a familiar location to most participants and contained several physical objects that could serve as scale references while still offering reasonable space to position the virtual objects and for the participants to move around them. Of course, people consume news articles in varied spaces, and future work should investigate the effect of varying environment sizes. 

\textbf{Participant demographics and sample size.} Our participant sample was reasonably large (26 participants) and included people with varying degrees of familiarity with AR. However, participants were recruited from a relatively homogeneous academic setting, with a mean age of 25 years---the 18-29 age group is known to be the one more accustomed to digital news \cite{pew2021news}. Our participants were recruited on the University campus, mainly in the buildings where computer science students have classes or research labs. Although this is usual in most VR/AR research, we acknowledge the limitations and bias that such a sample might induce.  Future studies should investigate whether similar findings extend to other demographic groups.

\section{Conclusion}
\label{sec:conclusion}

In this work, we have evaluated whether immersive infographics help people understand quantities expressed in news pieces through volumetric representations. We created a set of three immersive concrete scales infographics to test our hypotheses, which involved comparison with textual analogies and conventional image-based infographics. Results from our user study indicate that immersive infographics were significantly more effective than the text and image versions for the \textit{Trash} scenario and for all scenarios in combination, thus confirming our main hypothesis. These initial findings suggest that AR offers a suitable framework for journalists to help readers better understand the scale of the information conveyed in the news.

\section*{Acknowledgments}

We thank the participants of our experiments for allowing us to borrow their time and knowledge for this study, and the comments of the reviewers that helped us to improve the quality of this paper. We also acknowledge the support from the Brazilian National Council for Scientific and Technological Development (CNPq) and from Microsoft Research. This study was financed in part by the Coordenação de Aperfeiçoamento de Pessoal de Nível Superior - Brasil (CAPES) - Finance Code 001.

\renewcommand*{\bibfont}{\small}

\printbibliography

\end{document}